\documentclass[journal]{IEEEtran}

\usepackage[cmex10]{amsmath}
\usepackage{amssymb}
\usepackage{amsthm}
\usepackage{graphicx}
\usepackage{cite, color}
\usepackage{float}
\usepackage{multirow}
\usepackage{xfrac}
\usepackage{tikz}
\usetikzlibrary{shapes}
\usetikzlibrary{arrows.meta}
\usepackage{pgfplots}

\newtheorem{proposition}{Proposition}

\newcommand{\figurewidth}{1\columnwidth}
\newcommand{\figureheight}{0.8\columnwidth}

\begin{document}

\title{Design of LDPC Codes for the Unequal Power Two-User Gaussian Multiple Access Channel}%
\author{Alexios~Balatsoukas-Stimming and Athanasios~P.~Liavas%
\thanks{The authors are with the Electronic and Computer Engineering Department, Technical University of Crete,
73132 Chania, Greece (e-mail: alexios.balatsoukas@gmail.com, liavas@telecom.tuc.gr).}}

\maketitle

\begin{abstract}
In this work, we describe an LDPC code design framework for the unequal power two-user Gaussian multiple access channel
using EXIT charts. We show that the sum-rate of the LDPC codes designed using our proposed method can get close to the maximal sum-rate
of the two-user Gaussian multiple access channel. Moreover, we provide numerical simulation results that demonstrate the
excellent finite-length performance of the designed LDPC codes.
\end{abstract}

\section{Introduction}

\IEEEPARstart{I}{t} is known that corner points of the two-user Gaussian multiple access channel (GMAC) can be achieved
by successive decoding, while other points can be achieved by time-sharing, rate-splitting \cite{rimoldi1996}, or joint decoding
\cite{Amraoui2002}. The first work to provide an explicit low-complexity EXIT chart based method to design LDPC codes for the two-user GMAC under
joint decoding was~\cite{Roumy2007}. However, the authors of \cite{Roumy2007} considered only the case where
the two users have equal power. The problem was recently re-examined in \cite{Sharifi2016}, where a Gaussian
mixture (GM) model was used for the state-to-variable messages. This GM model was shown to improve the accuracy of the EXIT
chart based method of \cite{Roumy2007}, at the cost of increased code design complexity due to the need for an
expectation-maximization step at each iteration to estimate the parameters of the GM and because the resulting optimization problem is non-linear. The GM model was further improved
in \cite{Zheng2017} and used in conjunction with EXIT charts to design LDPC codes for the multi-user GMAC where all users
have equal power. Finally, the work of~\cite{Du2017} optimized the BPSK amplitudes used
to transmit groups of LDPC-coded bits under an equal average power constraint for the two-user GMAC.
Finally, a combination of IRA codes and repetition coding for the GMAC was examined in~\cite{Song2017}.

\subsubsection*{Contribution} In this work, we describe an LDPC code design method for the unequal power two-user GMAC,
which generalizes the method of~\cite{Roumy2007} and is a low-complexity alternative to~\cite{Sharifi2016}. As part of the
derivation of this method, we also provide a rigorous proof of the \emph{all-one/one-half codeword} assumption that has been
used throughout the relevant literature \cite{Roumy2007,Sharifi2016,Zheng2017,Du2017} without an explicit proof.

\section{Background}

\subsection{Two-User Gaussian Multiple Access Channel}
Let the length-$n$ binary codewords of the two users be $\mathbf{c}^{[j]} \in \mathcal{C}_j,~j=1,2,$ where $\mathcal{C}_j,~j=1,2,$ are the respective codes.
The BPSK modulated codewords are $\mathbf{x}^{[j]} = 1 - 2\mathbf{c}^{[j]},~j=1,2$. The output of the GMAC channel is
\begin{align}
\mathbf{y} & = \sqrt{P_1}\mathbf{x}^{[1]} + \sqrt{P_2}\mathbf{x}^{[2]} + \mathbf{w}, \quad \mathbf{w} \sim \mathcal{N}(0,\mathbf{I}_n),
\end{align}
where $P_1$ and $P_2$ denote the powers of the two users and, without loss of generality, the noise is assumed to have unit variance.
We define the design SNR for each user as $\text{SNR}_1 = P_1$ and $\text{SNR}_2 = P_2$. The following symmetry property of the GMAC can
be easily verified
\begin{align}
p({y_i}|x^{[1]}_i,x^{[2]}_i) & = p({-y_i}|-x^{[1]}_i,-x^{[2]}_i).
\label{eqn:unequalsym}
\end{align}

\subsection{LDPC Codes}

An ensemble of LDPC codes can be described by its edge perspective variable and check node degree distributions,
$\lambda (x)$ and $\rho (x)$, respectively, where \cite[p. 79]{Richardson2008}
\begin{align}
\lambda (x) & = \sum _{i} \lambda _i x^{i-1}, \qquad \rho (x) = \sum _{i} \rho _i x^{i-1}.
\end{align}
The node perspective variable node degree distribution $L(x)$ and the design rate R of the code are
\begin{align}
L(x) & = \sum _{i} L _i x^{i} = \frac{\int _0^x \lambda (z)dz}{\int _0^1 \lambda (z)dz}, \quad R = 1 - \frac{\sum _i \rho _i/i}{\sum _i \lambda _i/i}.
\end{align}

\subsection{Belief Propagation Decoding for the GMAC}

The belief propagation (BP) message-passing algorithm can be used to efficiently decode LDPC codes by exchanging log-likelihood ratio (LLR) messages
over the code's Tanner graph~\cite[Sec. 4.2]{Richardson2008}. If the Tanner graph is cycle-free, BP decoding is optimal with respect to the bit error rate.

Let $cv^{[j]}_{i}$ and $vc^{[j]}_{i}$ denote the check-to-variable
and variable-to-check messages for check and variable node $i$ of user $j$, respectively. These messages follow standard single-user BP rules~\cite[Sec. 3.3]{Richardson2008}.
Let $vs^{[j]}_{i}$ denote the variable-to-state message from user $j$ towards state node $i$ and let $sv^{[j]}_{i}$ denote the state-to-variable
message from state node ${i}$ towards user $j$. Messages $vs^{[j]}_{i}$ follow single-user BP rules~\cite[p. 59]{Richardson2008}. Using standard function node
message-passing rules~\cite[p. 56]{Richardson2008}, we can derive the following update rules for $sv^{[1]}_{i}$ and $sv^{[2]}_{i}$
\begin{align}
sv^{[1]} & = \log \frac{e^{-\frac{(y - \sqrt{P_1} - \sqrt{P_2})^2}{2}}e^{vs^{[2]}} + e^{-\frac{(y - \sqrt{P_1} +
\sqrt{P_2})^2}{2}}}{e^{-\frac{(y + \sqrt{P_1} - \sqrt{P_2})^2}{2}}e^{vs^{[2]}} + e^{-\frac{(y + \sqrt{P_1} + \sqrt{P_2})^2}{2}}},
\label{eqn:sv1}
\\
sv^{[2]} & = \log \frac{e^{-\frac{(y - \sqrt{P_1} - \sqrt{P_2})^2}{2}}e^{vs^{[1]}} + e^{-\frac{(y + \sqrt{P_1} -
\sqrt{P_2})^2}{2}}}{e^{-\frac{(y - \sqrt{P_1} + \sqrt{P_2})^2}{2}}e^{vs^{[1]}} + e^{-\frac{(y + \sqrt{P_1} + \sqrt{P_2})^2}{2}}},
\label{eqn:sv2}
\end{align}
where we have dropped the index $i$ for simplicity, since the update rules are identical for all state nodes. The above update rules are expansions of the general update rule of~\cite[Eq. (3)]{Roumy2007} for the unequal power case. The following symmetry properties of~\eqref{eqn:sv1} and~\eqref{eqn:sv2} can be easily verified
\begin{align}
sv^{[1]}({-y},{-vs^{[2]}}) & = -sv^{[1]}(y,vs^{[2]}), \label{eqn:sv1sym} \\
sv^{[2]}({-y},{-vs^{[1]}}) & = -sv^{[2]}(y,vs^{[1]}). \label{eqn:sv2sym}
\end{align}

\section{EXIT Charts for the Unequal Power GMAC}

In the limit of infinite blocklength, the Tanner graph of the two-user GMAC is cycle-free~\cite[p. 310]{Richardson2008}. Thus, we can use tools
such as density evolution (DE)~\cite{Richardson2001} and EXIT charts~\cite{tenbrink2004} to design LDPC codes for this channel.
Density evolution tracks the densities of the messages exchanged during BP decoding and can be used for the derivation
of conditions that guarantee a vanishingly small probability of error for large blocklengths. EXIT charts are a simpler analysis tool than DE that
reduces the infinite-dimensional problem of tracking densities to a single-dimensional problem of tracking the mutual information between the
messages in the decoder and the codeword bits.

\subsection{Restriction to the All-One/One-Half Codewords}

A crucial observation is that, for symmetric channels, the bit error probability of BP decoding is independent of the transmitted codeword \cite[Lemma 4.90]{Richardson2008}, meaning that the decoder analysis can be restricted to the all-one BPSK codeword, thus making the complexity of both DE and EXIT charts tractable.

Unfortunately, the GMAC is not symmetric with respect to each user. However, if we separately examine the
cases where $x^{[1]}_i=x^{[2]}_i$ (resp. $x^{[1]}_i = -x^{[2]}_i$), then the GMAC channel is equivalent to a BI-AWGN channel with inputs ${\pm\left( \sqrt{P_1}+\sqrt{P_2}\right)}$
(resp. ${\pm\left( \sqrt{P_1}-\sqrt{P_2} \right)}$), which is symmetric.
Let $a_1(y_i) = f(y_i|X_{i,1} = {+1})$ (resp. $a_2(y_i) = f(y_i|X_{i,2} = {+1})$) denote the density of the GMAC output $y_i$ conditioned on $X_{i,1} = {+1}$ (resp. $X_{i,2} = {+1}$),
where $X_{i,1}$ (resp. $X_{i,2}$) denotes the RV corresponding to $x_{i,1} = x^{[1]}_i = x^{[2]}_i$ (resp. $x_{i,2} = x^{[1]}_i = -x^{[2]}_i$). We can model each of the two BI-AWGN channels multiplicatively as~\cite{Richardson2001}
\begin{align}
  Y_{i,1} & = x_{i,1}Z_{i,1} \quad \text{and} \quad Y_{i,2} = x_{i,2}Z_{i,2}, \label{eqn:macmultsym}
\end{align}
where $Z_{i,1}$ and $Z_{i,2}$  are RVs that are distributed according to $a_{1}$ and $a_{2}$, respectively, and $x_{i,1}, x_{i,2} \in \{{-1},{+1}\}$.

An LDPC codeword is of type one-half if half of the codeword bits are equal to $0$ and half are equal to $1$~\cite[p. 296]{Richardson2008} (equivalently, if half of the BPSK modulated codeword bits are equal to ${+1}$ and half are equal to ${-1}$). LDPC codewords of type one-half are dominant for most codes as the blocklength goes to infinity~\cite[p. 517]{Richardson2008}, meaning that for the asymptotic analysis of the two-user GMAC, it is sufficient to consider the case where both users transmit a codeword of type one-half. The following proposition shows that the analysis can be further simplified, since for the two-user GMAC it is sufficient to consider the all-one BPSK codeword for one user and a type one-half codeword for the other user. In this case, half the state nodes will be $x^{[1]}_i=x^{[2]}_i$ nodes and half the state nodes will be $x^{[1]}_i=-x^{[2]}_i$ nodes.

\begin{proposition}
The DE analysis for the two-user GMAC can be restricted to the case where $\mathbf{x}^{[1]}$ is the
all-one BPSK codeword and $\mathbf{x}^{[2]}$ is a codeword of type one-half.\label{prop:1}
\end{proposition}
\begin{IEEEproof}
Similarly to \cite[Lemma 4.90]{Richardson2008}, we will show that
the probability of error when
the received values at $x^{[1]}_i=x^{[2]}_i$ nodes are
$\mathbf{Y}_1 = \mathbf{x}_1\mathbf{Z}_1$ and the received values at $x^{[1]}_i=-x^{[2]}_i$ nodes are $\mathbf{Y}_2 = \mathbf{x}_2\mathbf{Z}_2$
is equal to the probability of error for the case where the received values at $x^{[1]}_i=x^{[2]}_i$ nodes
are $\mathbf{Y}_1 = \mathbf{Z}_1$ and the received values at $x^{[1]}_i=-x^{[2]}_i$ nodes are $\mathbf{Y}_2 = \mathbf{Z}_2$.
Let $\mathbf{x} = [ \, \mathbf{x}_1 \;\, \mathbf{x}_2 \, ]$ denote the concatenation of $\mathbf{x}_1$ and $\mathbf{x}_2$ and let $x_i$ denote the $i$-th element of $\mathbf{x}$. In essence, we will show that the signs of the elements of $\mathbf{x}$ can be factored out in the message passing algorithm (like in the single-user case), and since bit-decisions only depend on the message signs, $\mathbf{x}$ can be restricted to the all-one BPSK codeword for the asymptotic analysis. This in turn implies, by construction of $\mathbf{x}$, that $\mathbf{x}^{[1]}$ can be the all-one BPSK codeword but $\mathbf{x}^{[2]}$ has to be a codeword of type one-half in order for the analysis to be valid.

Let $i_k$ be a variable node of
user $k$ and let $j_k$ be one of its neighboring check nodes. Let $vc _{i_kj_k}^{(\ell)}(y_{i})$ denote the message sent
from $i_k$ to $j_k$ in iteration $\ell$ assuming that the received value is $y_{i}$, let $cv _{j_ki_k}^{(\ell)}(y_{i})$ denote
the corresponding message sent from $j_k$ to $i_k$, and let $vs_{i_k}^{(\ell)}(y_{i})$ denote the message sent from variable node
$i_k$ to its corresponding state node. Finally, let $sv_{i_2}^{(\ell)}(y_i,vs^{(\ell)}_{i_1})$
(resp. $sv_{i_1}^{(\ell)}(y_i,vs^{(\ell)}_{i_2})$)
denote the message from the state node connected to variable node $i_1$ of user $1$ (resp. variable node $i_2$ of user $2$) towards the
corresponding variable node of user $2$, i.e. $i_2$ (resp. $i_1$).

\begin{figure*}[t]
  \centering
  \small
  \begin{align}
  F^{[1]}_{+}(\mu ) & = \frac{1}{\sqrt{\pi}}\int _{-\infty}^{+\infty}e^{-z^2} \log
  \left( \frac{ 1 + e^{\sqrt{4\mu + 8P_2}z + \mu + 2P_2}}{1 + e^{-\sqrt{4\mu + 8P_2}z - \mu - 2P_2 -4\sqrt{P_1}\sqrt{P_2}}}\right)dz - \mu + 2(P_1 - P_2)
  \label{eqn:mu1F1}\\
  F^{[1]}_{-}(\mu ) & = \frac{1}{\sqrt{\pi}}\int _{-\infty}^{+\infty}e^{-z^2} \log
  \left( \frac{1 + e^{-\sqrt{4\mu + 8P_2}z - \mu - 2P_2}}{1 + e^{\sqrt{4\mu + 8P_2}z + \mu + 2P_2 - 4\sqrt{P_1}\sqrt{P_2})}} \right)dz + \mu +
  2\left(\sqrt{P_1} - \sqrt{P_2}\right)^2
  \label{eqn:mu1F-1}\\
  F^{[2]}_{+}(\mu) & = \frac{1}{\sqrt{\pi}}\int _{-\infty}^{+\infty}e^{-z^2} \log
  \left( \frac{ 1 + e^{\sqrt{4\mu + 8P_1}z + \mu + 2P_1}}{1 + e^{-\sqrt{4\mu + 8P_1}z - \mu - 2P_1 - 4\sqrt{P_1}\sqrt{P_2}}}\right)dz - \mu + 2(P_2 - P_1)
  \label{eqn:mu2F1}\\
  F^{[2]}_{-}(\mu ) & = \frac{1}{\sqrt{\pi}}\int _{-\infty}^{+\infty}e^{-z^2} \log \left( \frac{1 + e^{\sqrt{4\mu + 8P_1}z +
  \mu + 2P_1 - 4\sqrt{P_1}\sqrt{P_2}} }{1 + e^{-\sqrt{4\mu + 8P_1}z - \mu - 2P_1}} \right)dz - \mu - 2\left(\sqrt{P_2} - \sqrt{P_1}\right)^2
  \label{eqn:mu2F-1}
  \end{align}\\
  \hrulefill
  \vspace{-0.25cm}
\end{figure*}

Due to \eqref{eqn:sv1sym} and \eqref{eqn:sv2sym}, the initial messages from the state nodes to the variable nodes of user $k$  are
\begin{align}
sv _{i_k}^{(0)}(y_{i},0) & = sv _{i_k}^{(0)}(x_{i}z_{i},0) = x_{i}sv _{i_k}^{(0)}(z_{i},0).
\end{align}
Due to the variable node update rule symmetry, we have
\begin{align}
vc _{i_kj_k}^{(0)}(y_{i}) & = vc _{i_kj_k}^{(0)}(x_{i}z_{i}) = x_{i}vc _{i_kj_k}^{(0)}(z_{i}).
\end{align}
Using the check and variable node symmetries \cite[Sec. 4.2]{Richardson2008}, we get
\begin{align}
vs _{i_k}^{(\ell+1)}(y_{i}) & = x_{i}vs _{i_k}^{(\ell+1)}(z_{i}).
\end{align}
Due to the state node update rule symmetry in \eqref{eqn:sv2sym}, for the message from the state node connected to variable
node $i_1$ towards the corresponding variable node of user $2$, we have
\begin{align}
sv_{i_2}^{(\ell+1)}(x_iz_i,x_{i}vs^{(\ell+1)}_{i_1}) & = x_{i}sv_{i_1}^{(\ell+1)}(z_i,vs^{(\ell+1)}_{i_1}).
\end{align}
An analogous statement holds for user $2$. By invoking the variable node symmetry again, we have
\begin{align}
vc _{i_kj_k}^{(\ell+1)}(y_{i}) & = x_{i}vc _{i_kj_k}^{(\ell+1)}(z_{i}).
\end{align}
As in \cite[Lemma 4.90]{Richardson2008}, since we can factor out the message signs at each iteration, we can conclude
that $\mathbf{x}$ can be the all-one BPSK codeword.
\end{IEEEproof}
\emph{Remark:} If we set $x_i = -x^{[1]}_i = x^{[2]}_i$ in (\ref{eqn:macmultsym}) when $x^{[1]}_i = -x^{[2]}_i$, then we will reach the conclusion that user $2$ can be restricted to
the all-zero codeword and user $1$ has to transmit a codeword of type one-half.

\subsection{Stability Condition \& Gaussian Approximation}
We observe that the receiver of an unequal power two-user MAC channel is equivalent to the best receiver of a two-user
broadcast channel \cite{berlin2005}. Thus, the stability condition is
\begin{align}
\lambda _2^{[j]}\sum_{i}(i-1)\rho _i ^{[j]} < \exp\left({P_j/2}\right),~j=1,2. \label{eqn:stability}
\end{align}
We note that the same stability condition was derived in \cite{Roumy2007} and \cite{Sharifi2016} using different arguments.

Under the Gaussian approximation (GA), all message densities are approximated as symmetric Gaussian and it is
sufficient to only track the means of these densities \cite{chung2001}.
For the unequal power two-user GMAC, the variable-to-check and check-to-variable messages follow standard GA rules.
The variable-to-state messages also follow standard GA rules with the difference that averaging is done over $L(x)$,
instead of $\lambda (x)$. Let us assume, without loss of generality, that the codeword of user $1$ is the all-one BPSK codeword,
while the codeword of user $2$ is a codeword of type one-half. If we further assume that the variable-to-state messages are
symmetric Gaussian with mean $\mu$ and variance $2\mu$, then we can derive the means of the
state-to-variable messages as in \eqref{eqn:mu1F1}--\eqref{eqn:mu2F-1}. The functions $F^{[1]}_{+}$ and $F^{[1]}_{-}$ (resp. $F^{[2]}_{+}$ and $F^{[2]}_{-}$)
are the means of the state-to-variable messages towards user $1$ (resp. user $2$) from state nodes that are connected
to a ${+1}$ and a ${-1}$ variable node of user $2$ (resp. user $1$), respectively. We note that these expressions are generalizations of
the expressions found in \cite{Roumy2007} and their detailed derivation can be found in~\cite{Balatsoukas2012}.

\subsection{EXIT Charts}
Let $I^{[j]}_{CV}$ (resp. $I^{[j]}_{SV}$) denote the mutual information between the codeword bits and the check-to-variable
(resp. state-to-variable) messages of user $j$. If we proceed as in~\cite{tenbrink2004}, we can show that the EXIT chart
$I^{i,[j]}_{VC}$ describing the variable-to-check messages for variable node of degree $i$ is
\begin{align}
  I^{i,[j]}_{VC} & = J\left(\sqrt{(i-1)\left[J^{-1}(I^{[j]}_{CV})\right]^2 + \left[J^{-1}(I^{[j]}_{SV})\right]^2}\right),
\end{align}
where the definition and
good approximations for $J(\cdot)$ and $J^{-1}(\cdot)$ can be found in \cite{tenbrink2004}.
Averaging over $\lambda (x)$, we get the variable-to-check EXIT chart
\begin{align}
  I^{[j]}_{VC} & = \sum _i \lambda _i I^{i,[j]}_{VC}. \label{eqn:evc1}
\end{align}
Similarly, it can be shown that the EXIT chart $I^{[j]}_{VS}$ describing the variable-to-state messages is
\begin{align}
  I^{[j]}_{VS} & = \sum _i L _i J\left(\sqrt{i}J^{-1}(I^{[j]}_{CV})\right).
\end{align}
By exploiting the duality between the check and variable nodes \cite[p. 236]{Richardson2008}, it can be shown that the EXIT
chart describing the check-to-variable messages can be well approximated as
\begin{align}
  I^{[j]}_{CV} & \approx \sum _i \rho _i\left[1 - J\left(\sqrt{(i-1)}J^{-1}(1-I^{[j]}_{VC})\right)\right],
\end{align}
where $I^{[j]}_{VC}$ is the mutual information between the variable-to-check messages and the codeword bits
and $I^{[j]}_{CV}$ is the overall check node EXIT chart for user $j$. Finally, the average mutual information between the state-to-variable messages towards user $1$ and this user's codeword bits is
\begin{align}
  I^{[1]}_{SV} & = \frac{1}{2}J\left( \sqrt{2F^{[1]}_{+}\left(\frac{1}{2}\left[J^{-1}\left(I^{[2]}_{VS}\right)\right]^2\right)} \right) \nonumber \\
& + \frac{1}{2}J\left( \sqrt{2F^{[1]}_{-}\left(\frac{1}{2} \left[J^{-1}\left(I^{[2]}_{VS}\right)\right]^2\right)} \right).
\end{align}
An analogous expression holds for user $2$~\cite{Balatsoukas2012}.

\begin{table*}[t]
  \centering
  \scriptsize
  \setlength{\tabcolsep}{2.75pt}
  \caption{LDPC Code Optimization Results.}\label{tab:results}
  \vspace{-0.1cm}
  \begin{tabular}{l|c|c|cccccccccccccccc}
                         & $C$                         & $R$        & $d_c$ & $\lambda_2$ & $\lambda_3$ & $\lambda_{11}$ & $\lambda_{12}$  & $\lambda_{13}$ & $\lambda_{22}$  & $\lambda_{23}$  & $\lambda_{28}$  & $\lambda_{29}$  & $\lambda_{34}$  & $\lambda_{35}$  & $\lambda_{98}$  & $\lambda_{99}$ &  $\lambda_{100}$ \\
    \hline
    User 1 $(P_1=1.5)$   &  \multirow{2}{*}{$0.886$}   & $0.505$    & $8$   & $0.2431$    & $0.3573$    &                &                 &                & $0.1511$        & $0.0745$        &                 &                 &                 &                 & $0.0412$        & $0.1328$       & \\
    User 2 $(P_2=1)$     &                             & $0.372$    & $7$   & $0.2248$    & $0.2990$    &                &                 & $0.1392$       &                 &                 & $0.0081$        & $0.0446$        &                 &                 &                 &                & $0.2843$ \\
    \hline
    User 1 $(P_1 = 3)$   &  \multirow{2}{*}{$1.115$}   & $0.726$    & $13$  & $0.2629$    & $0.4199$    &                &                 &                &                 &                 &                 &                 & $0.1291$        & $0.1881$        &                 &                & \\
    User 2 $(P_2=1)$     &                             & $0.370$    & $6$   & $0.2811$    & $0.3193$    & $0.0438$       & $0.1017$        &                &                 &                 &                 &                 & $0.1268$        & $0.0057$        &                 &                & $0.1216$ \\
    \hline
  \end{tabular}
  \vspace{-0.2cm}
\end{table*}

\section{Optimization and Results}

If the inverse of $I^{[j]}_{CV}$ lies below $I^{[j]}_{VC}$, for $j = 1,2$, then the probability of error of
BP decoding becomes vanishingly small~\cite{tenbrink2004}. For our code design procedure, we set the maximum variable node degree to $v_{\max}$.
Moreover, since $\rho(x) = x^{d_c-1}, \, d_c \in \mathbb{N}$,
lead to efficient single user designs \cite{chung2001}, we assume $\rho ^{[j]}(x) = x^{d_c^{[j]}-1}, d_c^{[j]} \in \mathbb{N}$.
Then, the inverse of $I^{[j]}_{CV}$, denoted by $I^{{-1},[j]}_{CV}$, can be approximated as
\begin{align}
  I^{{-1},[j]}_{CV} & \approx 1 - J\left( \frac{1}{\sqrt{d_c^{[j]}-1}}J^{-1}\left(1 - I^{[j]}_{CV}\right) \right).
\end{align}
Additionally, \eqref{eqn:stability} becomes $\lambda _2^{[j]} < \exp\left( {P_j/2}\right)/(d_c^{[j]}-1)$.
In general, each $I^{[j]}_{VC}$ is a non-linear function of the coefficients of $\lambda ^{[1]}(x)$ and $\lambda ^{[2]}(x)$, so the joint
code design can not be expressed as a linear program (LP). To overcome this problem, \cite{Roumy2007} uses the assumptions that
$\lambda ^{[1]}(x) = \lambda ^{[2]}(x)$ and that the state nodes always connect a degree $i$ variable node of user $1$ with a degree
$i$ variable node of user $2$. These assumptions are reasonable for the equal power case, but in the unequal power case the degree
distributions for each user have to be different in general to enable communication at different rates. In~\cite{Sharifi2016}, the authors
use differential evolution to optimize the degree distributions for the unequal power GMAC, which gives good results but it is less elegant
and has a much higher computational complexity than an LP approach.

In order to express the code design as an LP, we propose to fix the variable node degree distribution of one user and
optimize the variable node degree distribution of the other user by alternately solving the following LP, for $j=1,2$,
\begin{align}
  \text{maximize} \hspace{0.2cm} &  \sum _{i} \lambda ^{[j]}_i/i\\
  \text{subject to} \hspace{0.2cm} & I^{{-1},[j]}_{VC} < \sum _{i} \lambda ^{[j]} _i I^{i,[j]}_{VC}, \\
  & \sum _{i} \lambda ^{[j]}_i = 1,~\lambda ^{[j]}_i \geq 0, ~ i = 2,3,\hdots,v_{\max}, \\
  & \lambda ^{[j]}_2 < \exp\left( {P_j/2}\right)/(d_c^{[j]}-1).
\end{align}
The same procedure is repeated for several $(d_c^{[1]},d_c^{[2]})$ pairs and the best pair, in terms of sum-rate, is kept.

\begin{figure}
  \centering
  \begin{tikzpicture}

	\pgfplotsset{grid style={dashed}}
	\small

	\begin{semilogyaxis}[
		width = \figurewidth,
		height = \figureheight,
		xlabel = {Signal-to-Noise Ratio (dB)},
		ylabel = {Bit Error Rate},
		ylabel near ticks,
		xlabel near ticks,
		xmin = -0.025, xmax = 5.5,
		ymin = 5e-7, ymax = 10,
		grid = both,
		legend style={legend pos=north east,font=\scriptsize},
		legend cell align={left},
		legend columns={3},
		transpose legend,
	]

		\addlegendentry{\hspace{-0.7cm}$P_1 = 1.5$, $P_2 = 1\;$}
		\addlegendimage{empty legend}

		\addplot[blue, thick, solid, mark=*, mark options={scale=0.8}] table[x index=0, y index = 1] {figures/data/n_50000_P1_1.5_P2_1_user_1.dat};
		\addlegendentry{User $1$}
		\addplot[red, thick, dashed, mark=*, mark options={scale=0.8, solid}] table[x index=0, y index = 1] {figures/data/n_50000_P1_1.5_P2_1_user_2.dat};
		\addlegendentry{User $2$}

		\addlegendentry{\hspace{-0.7cm}$P_1 = 3$, $P_2 = 1$}
		\addlegendimage{empty legend}
		\addplot[blue, thick, solid, mark=square*, mark options={scale=0.8, solid}] table[x index=0, y index = 1] {figures/data/n_50000_P1_3_P2_1_user_1.dat};
		\addlegendentry{User $1$}
		\addplot[red, thick, dashed, mark=square*, mark options={scale=0.8, solid}] table[x index=0, y index = 1] {figures/data/n_50000_P1_3_P2_1_user_2.dat};
		\addlegendentry{User $2$}

    \addplot +[black, solid, ultra thick, mark=none] coordinates {(0, 5e-7) (0, 1e-4)};
    \addplot +[black, solid, ultra thick, mark=none] coordinates {(1.7609, 5e-7) (1.7609, 1e-4)};
    \addplot +[black, solid, ultra thick, mark=none] coordinates {(4.7712, 5e-7) (4.7712, 1e-4)};

		\node[anchor=west] (sourceP13) at (axis cs:2.85,1.5e-4){\scriptsize $\text{SNR}_1{=}4.77$~dB};
		\node (destinationP13) at (axis cs:4.75,9e-5){};
		\node[anchor=west] (sourceP115) at (axis cs:0.7,4e-4){\scriptsize $\text{SNR}_1{=}1.76$~dB};
		\node (destinationP115) at (axis cs:1.75,9e-5){};
		\node[anchor=west] (sourceP21) at (axis cs:0,1.3e-6){\scriptsize $\text{SNR}_2{=}0.00$~dB};
		\node (destinationP21) at (axis cs:0.04,1.1e-4){};

		\draw[-{Latex[length=1.75mm,width=1.25mm]}](sourceP13.east)--(destinationP13);
		\draw[-{Latex[length=1.75mm,width=1.25mm]}](sourceP115.south)--(destinationP115);
		\draw[-{Latex[length=1.75mm,width=1.25mm]}](sourceP21)--(destinationP21);

	\end{semilogyaxis}%
\end{tikzpicture}%
  \caption{Finite length performance of the optimized codes for $n = 50,000$.}
  \label{fig:macuneq}
  \vspace{-0.2cm}
\end{figure}
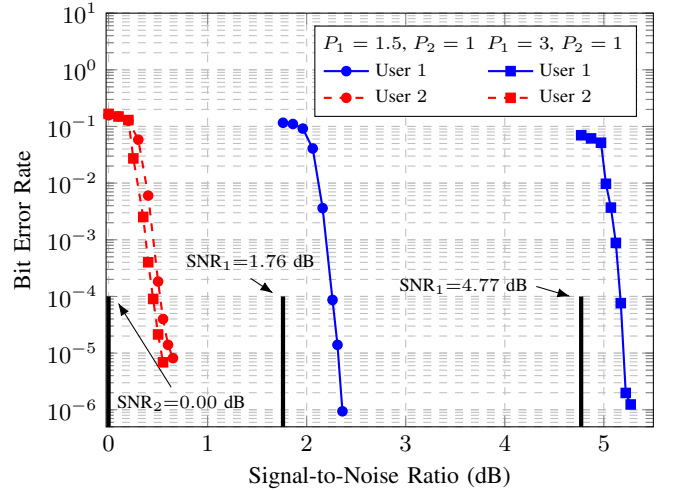

In order to test our method, we design LDPC codes for two unequal power cases ($P_1=1.5$, $P_2=1$ and $P_1=3$, $P_2=1$) and we set the maximum variable node degree to $v_{\max}=100$. The resulting degree distributions and corresponding rates are summarized in Table~\ref{tab:results}. We observe that, in both cases, the resulting codes have a sum-rate that is at most $0.02$ bits/(ch. use) away from the maximal sum-rate. Moreover, in Fig.~\ref{fig:macuneq} we show the finite-length performance of randomly constructed LDPC codes with no cycle removal and with $n = 50,000$ when performing $200$ decoding iterations. We observe that the designed codes achieve a BER of $10^{-5}$ at an SNR that is only approximately $0.6$~dB away from their design SNR. Finally, for $P_1=P_2$, our method gives identical results to~\cite{Roumy2007}.

\section{Conclusion}
In this paper, we presented an EXIT-chart based method for the design of LDPC codes for the unequal power two-user GMAC. To this end, we first showed that the asymptotic analysis can be restricted to the case where one user transmits the all-one BPSK codeword and the other user transmits a codeword of type one-half. We then showed that the code design problem can be expressed as an alternating sequence of LPs, which can be solved efficiently. Numerical results demonstrate that the resulting LDPC codes are at most $0.02$ bits/(ch. use) away from the maximal sum-rate and only $0.6$~dB away from their respective design SNRs when using randomly constructed LDPC codes of length $n = 50,000$.

\bibliographystyle{IEEEtran}
\bibliography{IEEEabrv,bibliography}

\end{document}